# Machine Learning Automated Approach for Enormous Synchrotron X-Ray Diffraction Data Interpretation

Xiaodong Zhao[1,2†], YiXuan Luo[3†], Juejing Liu[1,4†], Wenjun Liu[5], Kevin M. Rosso[1], Xiaofeng Guo[2,4*], Tong Geng[3*], Ang Li[1*], Xin Zhang[1*]

[1] Pacific Northwest National Laboratory, Richland, Washington 99354, United States

[2] Department of Chemistry, Washington State University, Pullman 99164

[3] Department of Electrical and Computer Engineering, University of Rochester, New York, 14627

[4] Materials Science and Engineering Program, Washington State University, Pullman 99164

[†] These authors contributed equally

* Corresponding Authors X.Z. (xin.zhang@pnnl.gov); A.L. (ang.li@pnnl.gov), T.G. (tgeng@UR.Rochester.edu), and X.G. (x.guo@wsu.edu)

**Abstract**:

Manual analysis of XRD data is usually laborious and time consuming. The deep neural network (DNN) based models trained by synthetic XRD patterns are proved to be an automatic, accurate, and high throughput method to analysis common XRD data collected from solid sample in ambient environment. However, it remains unknown that whether synthetic XRD based models are capable to solve μ-XRD mapping data for in-situ experiments involving liquid phase exhibiting lower quality with significant artifacts. In this study, we collected μ-XRD mapping data from an $LaCl_3$-calcite hydrothermal fluid system and trained two categories of models to solve the experimental XRD patterns. The models trained by synthetic XRD patterns show low accuracy (as low as 64%) when solving experimental μ-XRD mapping data. The accuracy of the DNN models was significantly improved (90% or above) when training them with the dataset containing both synthetic and small number of labeled experimental μ-XRD patterns. This study highlighted the importance of labeled experimental patterns on the training of DNN models to solve μ-XRD mapping data from in-situ experiments involving liquid phase.

Key words: Machine learning, Synchrotron XRD, Crystallization rare earth element, hydrothermal fluid.

## Introduction

X-ray diffraction (XRD) is a classical workhorse technique to study crystallography structures, such as crystal groups, phase compositions, unit cell parameters, and atomic displacement parameter.[1,2] The recently developed micro XRD (µ-XRD) method further expands the capability of XRD and brings the microscopic analysis to this classic technique.[3] The µ-XRD technique focuses the beam size of X-ray (300 nm in terms of diameter). Such small beam size is capable to reveal the spatial relationship among different phases in samples.[4-8] The combination of µ-XRD mapping and hydrothermal diamond anvil cell (HDAC) enables the study of such phase spatial relationships in the hydrothermal environment.[9-13]

However, the µ-XRD mapping generates significant amount of diffraction data. Traditionally, to qualitatively determine the phase information, e.g., identify and phase ratio from common XRD pattern, one must compare the current data with simulated or previously measured diffraction patterns. Such data interpretation requires high level of expertise, including coordination between data collection and analysis, skills to collect high quality data, recognition of systematic errors and preconception of potential phases. Even though there are commercial software, e.g., Jade,[14] for phase identification based on search and matching of previously reported XRD patterns, low precision and phases missing or mismatching often take place. In total, the manual analysis of XRD data requires labor-intensive manual checking and profound understanding to the materials. When using µ-XRD to study spatial relationships in samples, the amount of XRD data further scales up with an order of 2 making manual analysis of "big XRD data" challenging to the XRD community.[15-17]

A potential solution of such challenges is to use deep neural network (DNN) based models to extract information from the data by automatically recognizing features in the XRD patterns and hence tackling the "big data" problem.[18-24] By combining the 1D convolution layers and dense layers, DNN models are advantageous in extracting and learning features from labeled XRD patterns in training dataset. Multiple studies reported the adoption of DNN base models to extract various information from experimental XRD patterns, such as phase identify and ratio, unit cell parameter, and even distinguishing potential perovskite materials.[24-27] As DNN model training usually requires large amount of diverse data, a common practice is to generate theoretical XRD patterns from crystallography structures.[25-28] In previous studies, models trained by theoretical

XRD pattern is capable to analyze XRD spectra collected from solid samples in ambient environment.[24-27] However, it is remaining unknown that whether models trained by theoretical XRD pattern are capable to analysis data collected from µ-XRD mapping for in-situ experiment involving liquid phase.

Combining µ-XRD mapping and solid-liquid mixed samples in high-temperature and high-pressure environment is "the worst scenario" in terms of XRD data quality. As µ-XRD is usually adopted to analyze micro grains in the samples, common issues of resulting data may include overexposure, imperfect diffraction, and preferred orientation.[29] Although these issues create artefacts in the XRD pattern, the artefacts are manageable if the samples are totally solid.[4-8] Data with reasonable quality is still able to be retrieved. Involving liquid phase and extreme environments significantly amplify these adverse effects and distort the XRD data. Therefore, it pose a major challenge potentially preventing DNN models trained by synthetic XRD pattern from recognizing experimental µ-XRD data.

In this study, to verify whether the DNN models trained solely by synthetic XRD patterns are capable to analyze µ-XRD data collected from hydrothermal fluid environment, we trained multiple models to solve µ-XRD mapping data from the $LaCl_3$-calcite system at 200 °C. The accuracy of these models was evaluated by comparing the model driven results to the manual solving results. Two training datasets were generated, one only includes theoretical XRD patterns, while another is a combination of theoretical data and small number of labeled experimental patterns. Two types of models were trained, three binary classification models identifying the existence specified phases, and two multiclass multilabel models extracting all types and ratios of potential phases in the $LaCl_3$-calcite system. Our result shows that all DNN models trained by synthetic XRD data exhibit poor performance in terms of solving µ-XRD mapping data. Accurate and robust models was achieved only when small number of experimental XRD patterns were included into training dataset. This study highlighted the importance of labeled experimental data for DNN model training to solve µ-XRD mapping data collected from hydrothermal fluid systems.

## Methodology and Experiment

### Computation Platform

A conventional server cluster computer (processor: Intel Xeon Gold 6330 CPU, graphical processor unit: Nvidia A100-PCIE-40GB, memory: 40GB) was used to train all the models.

### Software Libraires

Software libraries used in this study were PyTorch, NumPy, Pandas, and SciPy. The python packages NumPy and Pandas were used to write code to load and preprocess the raw theoretical spectra (described below).[30, 31] PyTorch and TensorFlow2 were used to construct the NN and produce the models.[32-34] The code was written using Python 3.9.12.

### Data Augmentation and Preprocessing

Two categories of data were used to build training and evaluation datasets: theoretical XRD patterns and small number of labeled experimental data. The theoretical XRD patterns were generated by mixing two and three endmember patterns (bastnaesite, calcite, and Re metal). This mixing was conducted multiple times for every combination of phases with different ratios.[26-28] Small number of labeled experimental XRD spectra were also produced for generating datasets (see Additional information in SI for detailed composition of training datasets).

To unify the scale of theoretical data and experimental data, we extracted their spectrum intersection (2θ = 5° to 38°) and obtain the value over each 0.01° interval as the features, which means the number of features should be 3501. For the data that own random scale interval, we used 1D linear interpolation to fill in gaps. The diffraction amplitude is different between theoretical XRD pattern and experimental data. To ensure that the gradient moves smoothly towards the minima while maintaining the same rate for all the features, we scaled the features' value to [0,1] by data normalization as follows:

$$X = \frac{X - X_{min}}{X_{max} - X_{min}}, where\ X\ is\ the\ data\ sample.$$

The content above illustrated how we preprocessed the raw data for NN models. However, as a data-driven domain, machine learning requires numerous data samples, among which the experimental data are essential but hard to obtain by artificial labeling. Therefore, to enrich the existing unbalanced and limited dataset, we propose a data synthesis algorithm to generate artificial data points. In this data synthesis algorithm, $E$ and $T$ are the normalized experimental and theoretical datasets, respectively. By randomly selecting one positive sample $e_i$ from E and one negative sample $t_i$ from T, the artificially-created positive data sample can be represented by:

$$X_{ac} = e_i + \varepsilon \cdot t_i, where\ \varepsilon\ is\ a\ random\ number\ and\ \varepsilon \in (0,1).$$

Additionally, to unify the scale of original data and artificially create data, $X_{ac}$ also needs to be normalized using the approach mentioned above. Through the novel data synthesis approach, we can freely create a balanced and abundant dataset for training and validating, which greatly benefits the training process.

**NN Model Architecture.**

All DNN models trained in this study used multiple layers of convolution and maxpooling to extract the features from 1D XRD pattern. Based on the previously studies, the combination of these two layers improves the accuracy of models.[25-28] As shown in Figure S1, the binary classification DNN models determine the existence of phases from input XRD pattern mainly consists of 3 different kinds of layers including 1d convolution layer, maxpooling layer, and fully connected layer. At the start of the model, normalized data points are fed into a 1d convolution layer (input channel of 1, output channel of 4, kernel size of 5, stride size of 1, and padding size of 1) which effectively extracts the features from the input. Next, a maxpooling layer (kernel size of 16, stride size of 1, and padding size of 1) will process the output from the previous convolution layer, in which the large kernel size can enlarge the field of vision for peak detection. Then a combination of the convolution layer and the maxpooling layer with the same configuration will be added to the model. To make a prediction based on the features obtained from previous layers, we applied 6 fully connected layers to the model, among which the output nodes were set to be 1024, 512, 256, 128, 64, and 2. After forward propagation, the NN model will output 2 values denoting the class probabilities of the element, between which the class with a higher value will be considered as the classification result.

The architecture of the two multiclass and multi label models capable to retrieve all potential phases and phase ratio from input XRD pattern in the LaCl$_3$-calcite hydrothermal fluid system were obtained from a previous study (see Figure S2).[26] The datasets for training and evaluating the two models are similar to ones describe above, except the range of data is from 5.00° to 35.00° in terms of 2θ (2501 points). The only difference of the two models is that one was trained with dataset containing a small amount of labeled experimental data, and another by the dataset without any labeled experimental data.

**Training Notes.**

During the training phase for the binary classification models, we used CrossEntropyLoss (see Eq. 1) as the loss function which computes the cross-entropy loss between input logits and target.

$$Loss = -y_i \log(p(y_i)) + (1 - y_i) \cdot \log(1 - p(y_i)) \cdots (1)$$

To minimize the training loss, we used Adam, an algorithm for first-order gradient-based optimization of stochastic objective functions, to optimize the parameters in the models. With continuous iteration, Adam can adjust the parameters in our models to help them better fit observed data and determines the relation between input features and ground truth.

In addition to the choice of loss function and optimizer, we also need to set up other hyperparameters to initialize the training of models. To balance the trade-off between the rate of convergence and overshooting, we set the learning rate to be $1\times10^{-5}$. Furthermore, to prevent overfitting the model, the weight decay of the optimizer is set to be $1\times10^{-8}$. Besides, for the training and validation sets, the batch size was set to 60. Both training and evaluation dataset were shuffled at each training epoch. To ensure fairness while training, we set the training epoch to 200 for each element.

For the multiclass and multilabel models, we modified the cross entropy loss function and accuracy metric function based on the previous study.[26] Adam optimizer was used as the gradient decent function with the initial learning rate as $1\times10^{-2}$ and gradually decrease to $1\times10^{-8}$. Similar to the binary classification models above, the training and evaluation dataset were shuffled during each epoch. These models were trained by 256 epochs.

**Training and evaluation of different NN models.**

Even though we artificially created data for training and validation, the samples in the test set were all from experimental data collected from the LaCl$_3$-calcite hydrothermal fluid system, which means the test set is still unbalanced. While evaluating a dataset owning many more negative samples than positive samples, accuracy is not enough to fairly demonstrated model performance.

$$Acc = \frac{Number\ of\ correct\ predictions}{Total\ number\ of\ predictions}$$

Hence, in addition to the accuracy, we also applied the area under the receiver operating characteristic (AUROC) and the area under the precision-recall curve (AURPC) to prove the superiority of our model (see Figure S3 and additional information in SI for detailed algorism).

The evaluation of the two comprehensive models were performed by a direct comparison method. The model driven spatial distribution of different phases was compared with the manual solving results.

**Theoretical XRD pattern generation.**

The three main features of XRD patterns we considered in this work are: *i*) peak position, *ii*) peak intensity; *iii*) peak profile shape. A Python script was used to manipulate General Structure Analysis System-II (GSAS-II) to generate theoretical XRD patterns without use of the graphic user interface. GSAS-II is a software package developed by Toby, B. H., & Von Dreele, R. B. for X-ray and neutron diffraction analysis.[35] For the same compound, the three features of XRD can be impacted by the measuring condition, experiment where it is performed, source of X-ray, and sample nature.

The principle of XRD method is based on the diffraction of X-rays by crystalline structure, with the diffraction rule described by Bragg's law,

$$n\lambda = 2d_{hkl} \sin(\theta).$$

The d spacing is defined by the unit cell parameters (a, b, c, α, β, γ) and miller indices (h, k, l).

$$\frac{1}{d_k^2} = \frac{1}{v^2}\begin{bmatrix} H^2b^2c^2 \sin^2\alpha + K^2a^2c^2 \sin^2\beta + L^2a^2b^2 \sin^2\gamma + 2HKabc^2(\cos\alpha\cos\beta - \cos\gamma) + \\ 2KLa^2bc(\cos\beta\cos\gamma - \cos\alpha) + 2HLab^2c(\cos\gamma\cos\alpha - \cos\beta) \end{bmatrix},$$

where

$$v = abc(1 + 2\cos\alpha\cos\beta\cos\gamma - \cos^2\alpha - \cos^2\beta - \cos^2\gamma)^{\frac{1}{2}}.$$

In order to change the d spacing, the unit cell parameters a, b and c of different crystalline phases were modified randomly from 1% to 10%, so the peak position of the theoretical XRD patterns were modulated accordingly.

The diffracted intensities $I_{(hkl)}$ are proportional to the square of structural factor $F_{(hkl)}$,

$$F_{(hkl)} = \sum_{j=1}^{N} f_j \times \exp\left(-i2\pi(hx_j + ky_j + lz_j)\right).$$

Where $f_j$ is the atomic form factor for element j, hkl is the miller indices and (x, y, z) is the coordinate of atoms of element j in the unit cell.

While the total intensity $I_{(hkl)}$ is formulated as low:

$$I_{(hkl)} = K \times |F_{(hkl)}|^2 \times f_a \exp\frac{-B\sin^2(\theta)}{\lambda^2} \times A \times L(\theta) \times P(\theta) \times m.$$

Where K is a constant, $f_a \exp\frac{-B\sin^2(\theta)}{\lambda^2}$ stands for the thermal displacement off the equilibrium position due to temperature effect, A is the absorption factor,

The peak shape function $G_k$ is a key component to simulate a XRD pattern. The XRD pattern profile is generally controlled by instrument factors and sample factors. The instrument effect can be approximated by Gaussian function due to the similarity in the peak shape, and the sample effect can be reproduced by Lorentzian function. By modulating these factors through the Pseudo-Voigt and Pearson VII function, a XRD pattern can be easily synthesized. In this work, the pseudo-Voigt regime was used to simulate the profile function of the XRD peaks by tuning the Gaussian (G) and Lorentzian (L) components. The Gaussian shape is elucidated by the Cagliotti function

$$G \approx H_k^2 \approx U \times \tan^2\theta + V \times \tan\theta + W + \frac{P}{\cos^2\theta}.$$

where U, V, W and P parameters were used to control he peak profile.

Then other important factors, including the effect from crystallite size broadening, strain broadening, are controlled by the Lorentzian terms X and Y. Overall, the Pseudo-Voigt function is a linear combination of Gaussian function and Lorentzian function by the ratio of

$$pV(x) = \eta G(x) + (1 - \eta)L(x).$$

Above all, the peak center (position of maximum), height (height of the peak at the maximum) and FWHM (full width at half maximum of the peak) of theoretical XRD patterns were achieved by modifying the unit cell parameters (a, b, c), Gaussian profile parameters (U, V, W), and Lorentzian parameters (X, Y).

**Experimental XRD acquisition and preprocessing.**

Synchrotron X-ray diffraction were performed at the APS. A microdiffraction µ-XRD technique was employed to investigate the spatial correlation in the camber at the sector 34-ID-B. The X-ray beam size was 300 nm and X-ray energy was at 22 keV, enabling the spatial resolution (see Figure S4). 110 µ-XRD patterns were collected at temperature of 200 °C. Less data was collected under higher temperatures due to shorter data acquisition time used to prevent liquid leakage during heating. Collected 2D diffraction images were calibrated, masked, and integrated by Dioptas software and General Structure Analysis System software version II (GSAS-II) software. [35, 36] The background of 1D patterns was subtracted through Dioptas software, the polynomial order was up to $50^{th}$ order with a smoothing width at 0.1 Å, and the iteration was 150 times. The concentration of $LaCl_3$ for collecting µ-XRD from the $LaCl_3$-calcite hydrothermal fluid system was 0.1M with calcite. The temperature was set to 200 °C.

## Results and Discussion

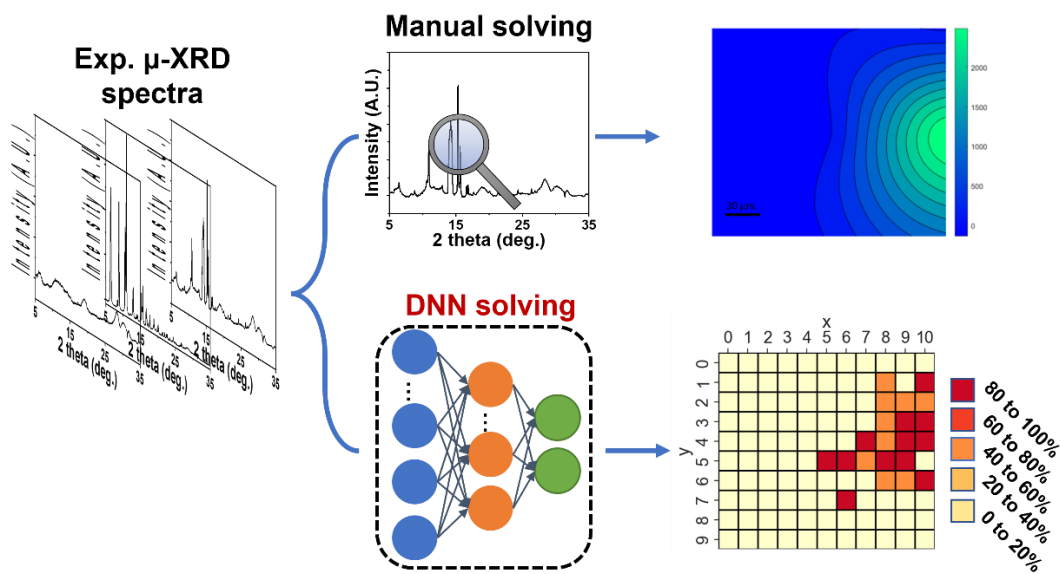

**Figure 1.** Illustration of solving μ-XRD mapping data by traditional search-and-match method and deep neural network based automatic method.

XRD mapping generates significant amount of data (see Figure S4). A high throughput automatic analysis method is therefore in needed to process the μ-XRD data and retrieve phase type and ratio information. As shown in Figure 1, we collected the μ-XRD mapping data from the $LaCl_3$-calcite hydrothermal fluid system with specified spatial configuration (a matrix with 11 columns and 10 rows, 110 XRD patterns in total) to observe the relationships among different mineral phases. We then used two methods to analyze the type and ratio of phases in every XRD pattern, manual mapping and an DNN model-based automatic analysis. When training model by a dataset with both theoretical XRD patterns and small number of labeled XRD patterns, the phase information driven from DNN model matches well with the manual analysis method.

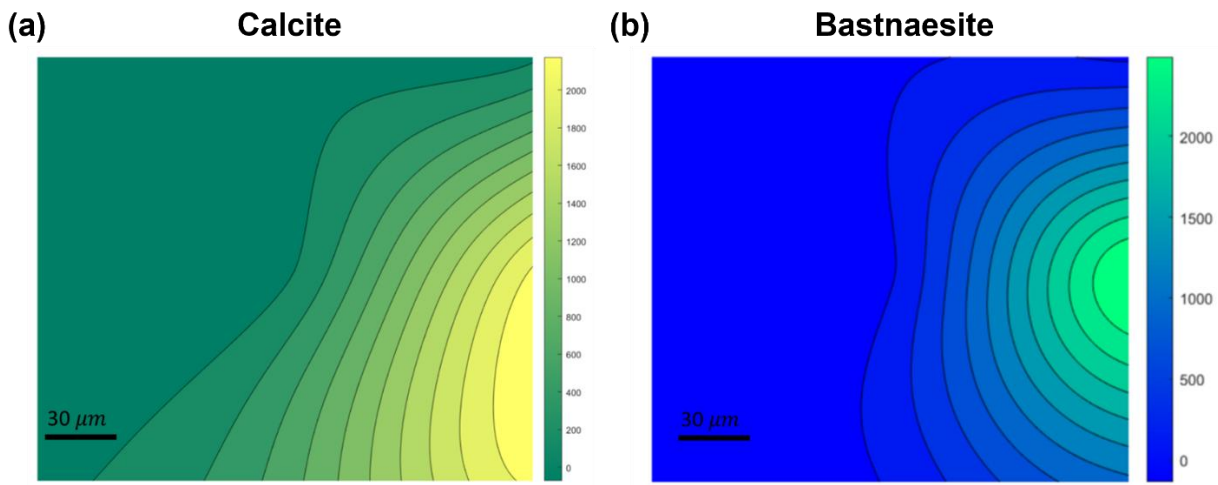

**Figure 2**. (a) Heatmap of calcite occupancy in the LaCl$_3$-calcite hydrothermal fluid system at 200 °C. Bright yellow color represents a high enrichment of calcite in the area. (b) Heatmap of La-bastnaesite occupancy in LaCl3-calcite system at 200 °C. Bright cyan color represents a high enrichment of calcite in the area.

A µ-XRD mapping result collected from the LaCl$_3$-calcite hydrothermal fluid system at 200 °C shows an overall overlapped spatial relation between calcite and La-bastnaesite (see Figure 2). In this experiment, 110 XRD patterns were collected with a $11 \times 10$ matrix. The phase quantity is determined by the characteristic peak intensity within a width of 0.1° signals from 8.80° to 8.90° all counted for bastnaesite and signals from 10.55° to 10.65° for calcite. The phase quantity is represented by the color brightness, with brighter color suggesting higher. It is illustrated that calcite phase is centered in the right bottom corner of the scanned zone, which is coherent with the X-ray imaging shown in Figure S4. This is because the calcite solid was initially introduced to that position, and La$^{3+}$ existed as an ion form in the solution. With dissolution of calcite at 200 °C, the released carbonate ion facilitates the precipitation of La$^{3+}$ in the form of bastnaesite (LaCO$_3$OH) solid form. It is worth noting that the center of bastnaesite precipitation zone is by the edge of calcite enriched area, which indicates a spatial dependence of La-bastnaesite mineralization in conjunction with the presence of calcite. Re metal from the HDAC was also found at the bottom site. It should be noticed that these heatmaps are obtained by manually analyzing 110 XRD patterns, while only corresponding with one experimental parameter (the LaCl$_3$-calcite hydrothermal fluid system with 0.1 M of $La^{3+}$ at 200 °C). Therefore, it is of importance to develop an automatic method to analyze the future µ-XRD mapping data from REE hydrothermal fluid system.

**Table 1**. Performance of DNN models in terms of phase identification trained by datasets with and without experimental data (w/ exp. and w/o exp.). See Additional information in SI for detailed composition of training datasets.

| Phase | AUROC | | AUPRC | | Accuracy (%) | |
|---|---|---|---|---|---|---|
| | w/o exp. | w/ exp. | w/o exp. | w/ exp. | w/o exp. | w/ exp. |
| Bastnaesite | 0.96 | 0.96 | 0.88 | 0.92 | 89 | 92 |
| Calcite | 0.66 | 0.95 | 0.62 | 0.88 | 64 | 90 |
| Re | 0.50 | 0.96 | 0.26 | 0.99 | 74 | 95 |

As a proof-of-concept, we first trained three deep neural networks (DNN) models (bastnaesite, calcite, and Re metal) to analyze the experimental diffraction data. These binary classification models exclusively identify the corresponding phases and output true or false when the phase is present or absent in the input XRD pattern, respectively. Two kinds of datasets were used to train these models. The dataset without labeled experimental data was generated by a previous reported method.[26] Another dataset was obtained by adding a small number of labeled experimental diffraction patterns into the first dataset. To verify the impact of different training dataset on the performance of models, we used these models to analyze experimental data manually solved (see Figure 2) but not used in training or evaluation of models. Three metrics were used to evaluate the performance, area under the receiver operating characteristic (AUROC), area under precision-recall curve (AUPRC), and accuracy (see method section for detailed information).

Comparing performance of DNN using and without using experimental data inputs for training (Table 1) shows that solely using synthetic diffraction patterns is not sufficient to train models that can identify phases from our experimental data, although this method has previously been reported in multiple studies.[25-28, 37, 38] As shown in Table 1, most models trained by dataset without labeled experimental data shows lower performance in terms of phase identification, including lower values of AUROC, AUPRC and accuracy. The only exception is bastnaesite focused models. The AUROCs of two bastnaesite focused models are the same, 0.96. As the AUROCs are closed to

1.00, the models are confident about judging whether the experimental data contains bastnaesite. For the AUPRC and accuracy, the model trained without experimental data is slightly lower than that trained with experimental data. Overall, the two bastnaesite focused models behave similar in terms of deciding existence of bastnaesite phase from experimental data plausibly due to the XRD patters of bastnaesite phase is significantly different than the calcite and Re metal phases (see Figure S5).

For the rest of the two models, the difference of performance is large between models trained with and without experimental data. The AUROC and AUPRC of calcite focused model trained without experimental data is only 0.66 and 0.62, which are significantly lower than 0.95 and 0.88 from model trained with experimental data. The accuracy of model trained without experimental data is only 64%, closed to random guessing (50%). In contrast, the accuracy of model trained with experimental data is 90%. Similar results can be seen in Re metal focused model. Although the accuracy of the two models trained without experimental data are better than the calcite one (74% for Re), the very low AUROC and AUPRC indicated that the robustness of these models is weak. Therefore, the models trained without experimental data are not capable to identify phases from experimental data.

The low performance of models trained without experimental data is due to the big differences between synthetic XRD pattern used in training and µ-XRD data collected from experiment. In previous studies, the experimental data for evaluation of DNN models was usually collected from solid samples at ambient environment.[26-28] High quality X-ray diffraction peaks from all different planes were easy to obtain by the instrument. The resulting diffraction pattern can then be fitted easily with the synthetic data generated from crystallography structure. In contrast, the µ-XRD data collected from hydrothermal fluid system only show a few primary diffraction peaks in the best scenario due to several adverse effects including the small beam size, low exposure time/flux, poorly crystallinity of sample, preferred orientation, and overexposure in 2D images (see Figure S6). The differences between the theoretical and experimental XRD patterns, including intensity distortion of primary peaks and missing of many secondary peaks, are beyond what the model if only trained by theoretical data can handle, even both corresponding to the to the same crystal structure. Therefore, adding labeled experimental data into training dataset is crucial to improve the robustness of the DNN-based XRD analysis models to the data collected from hydrothermal fluid system.

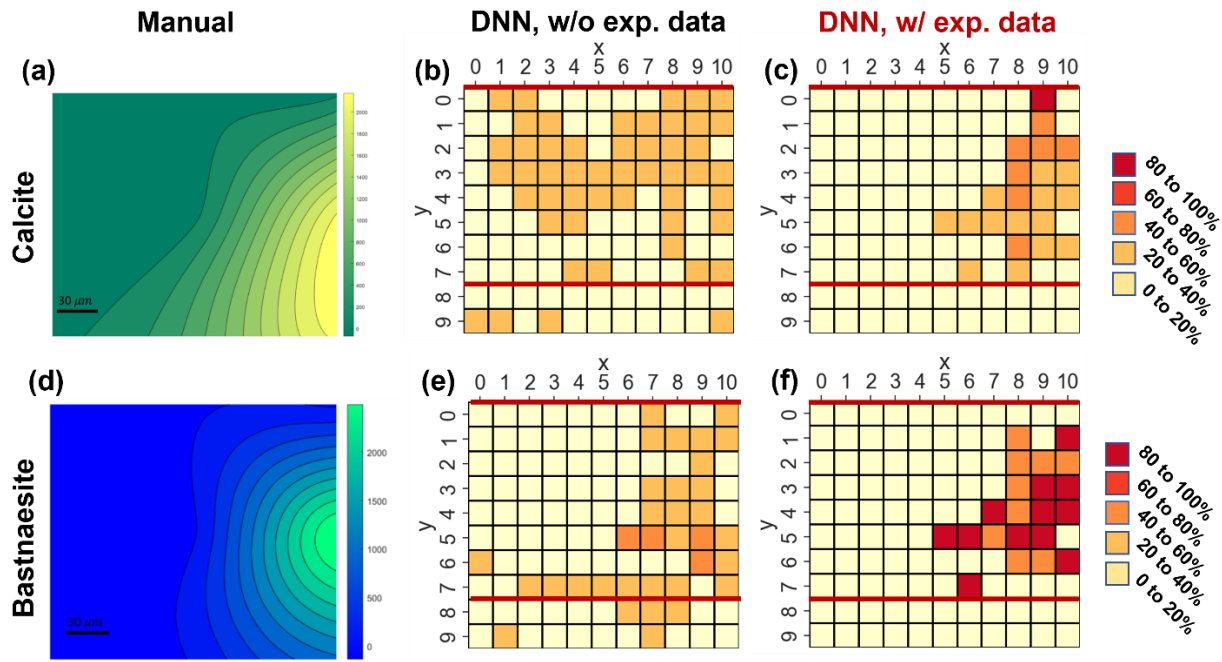

Figure 3. Utilization of DNN based models to retrieve phase types and ratios from μ-XRD mapping data obtained from the LaCl$_3$-calcite hydrothermal fluid system at 200 °C. (a) to (c) Distribution of calcite obtained by manually mapping (a), DNN model trained without experimental data (b), and DNN model trained with experimental data (c). (d) to (f) Distribution of bastnaesite obtained by manually mapping (d), DNN model trained without experimental data (e), and DNN model trained with experimental data (f). The read lines in the heatmaps retrieved by DNN based models mark the region shown in manual analysis.

We further trained two multiclass and multilabel DNN models (see Figure S2 and methodology section for model architecture) to not only identify the existence of all possible phases in the LaCl$_3$-calcite system but also ratio of phases from experimental XRD patterns. The model trained with dataset containing small number of labeled XRD pattern exhibits considerably better accuracy than that trained without labeled experimental data. As shown in Figure 3a and b, there is a big difference in the spatial distributions of calcite from manual solving vs. DNN models trained without experimental data. The manual solving result shows that the calcite conjugates at the right side of the area. In contrast, the DNN models trained without experimental data suggests that calcite is virtually everywhere. After adding a small amount of labeled experimental data, the model successfully retrieves the distribution of calcite in the area (see Figure 3c), which is closed to that from the manual solved.

A similar trend is also seen in the retrieving of bastnaesite from experimental data (see Figure 3d, e, and f). For the result based on model trained without experimental data, although it roughly retrieved the distribution of bastnaesite in the area, it falsely predicts bastnaesite at the bottom of the region of interest (right above the red line). Adding experimental data significantly improved the performance of the model resulting in the retrieval of bastnaesite phase distribution similar to the manual solved. Moreover, the model trained with experimental data is less likely to falsely predict phases not seen in the manual solved results in the $LaCl_3$-calcite hydrothermal fluid system, such as $La_2O_2CO_3$ with *Ama2* or *C12c1* space group and $LaOHCO_3$ (see Figure S7 and S8). In total, these two models further present the importance of the experimental data in training dataset when utilizing DNN to automatically solve XRD data from hydrothermal fluid system.

**Conclusion**

In this study, we demonstrated the importance experimental data in training dataset when training DNN models to obtain phase information from µ-XRD mapping data collected from hydrothermal fluid systems. The simple binary model experiment shows that model trained with small amount of experimental data outperforms the models trained without experimental data when judging the existence of phases based on experimental µ-XRD data. All three statical parameters (AUROC, AUPRC, and accuracy) of models trained with experimental data are higher/better that those trained without experimental data. This trend is maintained when using the same datasets to train two more comprehensive models to retrieve not only type but also ratio of phases. The model trained with experimental dataset correctly retrieved the spatial distribution of calcite and bastnaesite from µ-XRD mapping data. This study emphasized that training DNN models with synthetic XRD patterns is not the universal solution to analyze all XRD data. It also highlighted the necessity to build an experimental REE mineral dataset in the extreme environment, which would be great of importance for the future DNN studies trying to identify REE mineral from hydrothermal fluid environment.

**Acknowledgement:**

This research was supported by a Laboratory Directed Research and Development Project at Pacific Northwest National Laboratory (PNNL). A portion of the work was performed using the Environmental and Molecular Sciences Laboratory (EMSL), a national scientific user facility at PNNL sponsored by the DOE's Office of Biological and Environmental Research. PNNL is a multi-program national laboratory operated by Battelle Memorial Institute under contract no. DE-AC05-76RL01830 for the DOE. This research used resources of the Advanced Photon Source, a U.S. Department of Energy (DOE) Office of Science user facility operated for the DOE Office of Science by Argonne National Laboratory under Contract No. DE-AC02-06CH11357. X.Z., J.L., and X.G. also acknowledge the support of this work by the National Science Foundation (NSF), Division of Earth Sciences, under award No. 2149848.